\documentclass[twocolumn,linenumbers]{aastex631}

\usepackage{newtxtext,newtxmath}
\usepackage{verbatim}
\usepackage{color}
\usepackage{makecell}
\usepackage{enumerate}
\usepackage{array}

\usepackage[flushleft]{threeparttable}

\begin{document}

\title{The spin measurement of the black hole SLX 1746-331 using \textit{Insight}-HXMT observations}


\correspondingauthor{Wei Wang}
\email{wangwei2017@whu.edu.cn}

\author[0000-0001-5514-1167]{Jiashi Chen}
\affiliation{Department of Astronomy, School of Physics and Technology,
Wuhan University, Wuhan 430072, China}

\author[0000-0003-3901-8403]{Wei Wang}
\affiliation{Department of Astronomy, School of Physics and Technology, Wuhan University, Wuhan 430072, China}
 
\begin{abstract}
We present an X-ray spectral analysis of a black hole X-ray binary SLX 1746-331 during the 2023 outburst using five \textit{Insight}-HXMT observations. We jointly use the reflection model \texttt{relxillcp} and the general relativistic thermal thin disk model \texttt{kerrbb} to fit the spectra from 2 - 100 keV. By jointly fitting the five spectra, we constrained the black hole mass to be $\rm{M}=\rm{5.8\pm0.3~M_{\sun}}$ and dimensionless spin parameter to be $a_{*}=0.88^{+0.01}_{-0.02}$ (90 percent statistical confidence). The reflection model shows that SLX 1746-331 is a high-inclination system with the inclination angle $i=63.7^{+1.3}_{-1.0}$ degrees, the accretion disk has a density $\rm{log}N\sim 16 ~\rm cm^{-3}$. In addition, with the different reflection model \texttt{relxilllp}, which assumes a lamp-post geometry corona, we still give similar results.

\end{abstract}

\keywords{Black hole physics --- X-rays: binaries --- accretion, accretion disks}

\section{Introduction}

Outbursts of X-ray binaries are believed to be powered by matter accretion onto stellar-mass black holes or neutron stars. The matter in the accretion disk falls into a black hole (BH) releasing its gravitational potential energy and emitting thermal radiation in the UV/X-ray bands \citep{1973A&A....24..337S}. The accretion disk may evaporate and float into a large-scale height and play the role of the corona. The power-law component is generally thought to originate from Compton up-scattering of soft X-ray photons from the disk by a cloud of hot electrons (corona) located close to the black hole \citep{1979Natur.279..506S}. The spectrum may also display a reflection component that is coming from the accretion disk reflecting the corona emission. The reflected emission will be reprocessed by the disk's upper atmosphere and re-emit with characteristic features such as an iron K$\alpha$ fluorescence line at $\sim$ 6.4 - 6.97 keV, depending on the ionization of the iron ions, and a broad bump peaking at $\sim$ 20 - 30 keV referred to as the Compton hump \citep{2010ApJ...718..695G}. The reflection features near the BH will be distorted by the relativistic motion of disk material and the gravitational redshift. Therefore, the profiles of reflection features are directly linked to the inner radius of the accretion disk. Thus, the analysis of reflection features can be used to study accreting black holes, investigate the properties of the accretion disks, measure black hole spins, and even test Einstein’s theory of general relativity in the strong field regime \citep{1972ApJ...178..347B,2014SSRv..183..277R,2017RvMP...89b5001B}.

Spin is a fundamental physical parameter of a black hole, which has a significant effect on the accretion process. The efficiency of accreting matter converted to radiation is sensitive to the black hole spin and can vary up to an order of magnitude depending on whether the matter accretes onto a slowly or rapidly rotating black hole \citep{1972ApJ...178..347B}. The kinetic luminosity of relativistic jets produced by a black hole is probably tied to its spin state. However, the ingredients necessary to generate jets are poorly understood. The state of accretion flow and the rotation of the accretion disk will also influence the generation and properties of jet \citep{2020ARA&A..58..407D}. Spin can help us understand stellar-mass black holes in Galactic X-ray binary systems. In such systems, the spin can be a window on the formation process of black holes which mostly formed from the core collapse of massive stars \citep{2021ARA&A..59..117R}. The spin is commonly defined in terms of the dimensionless parameter $a_{*} \equiv a/M = cJ/GM^2$, where $M$ and $J$ represent the black hole mass and angular momentum, respectively.

There are two prevailing methods for measuring the black hole spin: the continuum-fitting method \citep{1997ApJ...482L.155Z,2014SSRv..183..295M,2021ApJ...916..108Z} and the X-ray reflection method \citep{1989MNRAS.238..729F,2006ApJ...652.1028B}. Both methods are based on the assumption that the accretion disk is a geometrically thin, optically thick, and innermost stable circular orbit (ISCO) effectively truncates the observable disk. The thermal continuum fitting (CF) method is based on the fact that the spin of a black hole influences the position of ISCO as well as the temperature of the inner disk. For higher spin, the smaller ISCO means that more binding energy is extracted from accretion matter and heats the inner disk to a higher temperature \citep{2021ARA&A..59..117R}. This method relies on accurate measurement of the system parameters of mass, distance, and inclination angle for the source. These parameters can be reliably measured via independent methods \citep{2009ApJ...706L.230M, 2001AJ....122.2668G, 1998ApJ...499..375O}. The CF method is applied more widely in black hole X-ray binaries than AGNs, because the thermal spectra of AGNs is in the UV band and the galactic absorption limits the possibility of accurate measurements.

\begin{figure}
	\includegraphics[width=\columnwidth]{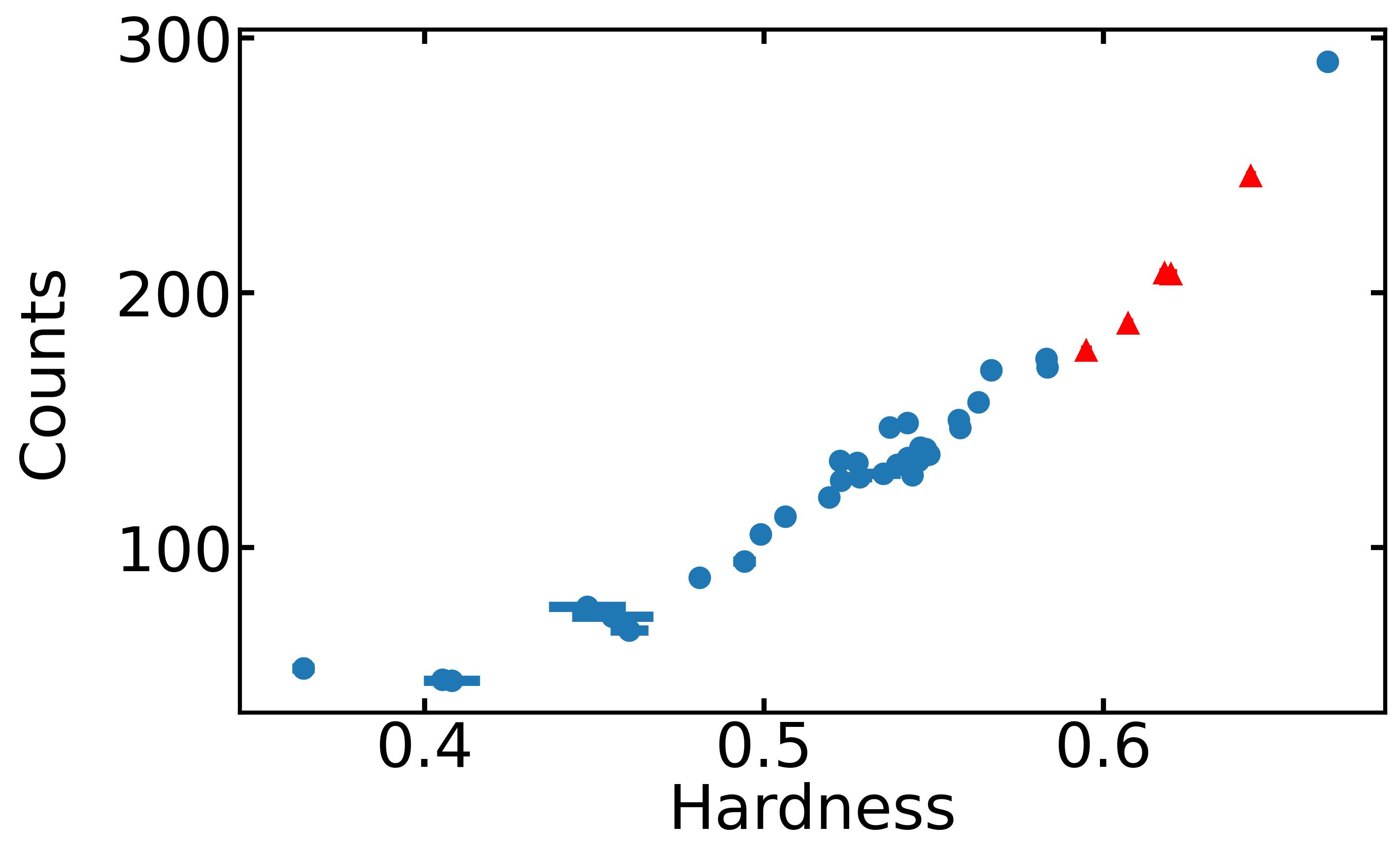}
    \caption{The evolutionary tracks in hardness-intensity diagram for \textit{Insight}-HXMT observations. The vertical axis presents the count rate in energy band 2 - 10 keV. The horizontal axis presents the hardness ratio (HR) defined as the ratio of count rate between 2 - 4 keV and 4 - 10 keV. The red triangles represent the Obs. $1-5$.}
    \label{figure1}
\end{figure}

\begin{table}
    \centering
    \caption{The list of HXMT observation IDs of the source SLX 1746-331 considered for the study.}
    \begin{tabular}{c c c}
       \hline
       Num. & Observation ID & MJD\\
       \hline
       1 & P051436300202  &   60018  \\
       2 & P051436300302  &   60020  \\
       3 & P051436300501  &   60021  \\
       4 & P051436300701  &   60022  \\
       5 & P051436300802  &   60023  \\
       \hline
    \end{tabular}
    \label{table1}
\end{table}

In the X-ray reflection method, spin is measured from the gravitational redshift of spectral features (fluorescent lines that can be determined by atomic physics, absorption edges and recombination continua) close to the ISCO. This method can be well applied to measure the spin of both supermassive BHs in AGNs and stellar mass BHs in X-ray binaries. In addition, it can make an independent constraint on the disk inclination angle via its effect on the shape of spectral lines \citep{2014ApJ...793L..33M, 2020MNRAS.493.4409D}. In addition to the above two methods, gravitational wave (GW) astronomy has opened a new window on black hole spin via the study of merging binary BHs \citep{2021ARA&A..59..117R}. GW signatures of relativistic gravity, including spin, are “clean” in the sense that they are not subject to the complexities affecting our understanding of accretion flows. However, the imprints of spin on GWs can be subtle and, at the current level of sensitivity provided by the Laser Interferometer Gravitational-Wave Observatory (LIGO-Virgo-KAGRA experiment), there are still only a small number of strong spin constraints \citep{2019PhRvX...9c1040A}.

SLX 1746-331 is a transient low-mass X-ray binary located at the Galactic center. The transient black hole X-ray binaries typically remain in a long-term quiescent state with a low accretion rate. As the accreted matter in the disk accumulates and releases gravitational potential, the disk will reach a high temperature, and hydrogen in it will be ionized. Due to thermal and viscous instabilities, the angular momentum transfers outward as the material falls inward, the accretion rate increases and an X-ray outburst will start \citep{1995ApJ...454..880C,2001NewAR..45..449L,2011BASI...39..409B}. SLX 1746-331 was identified within the surveyed fields of the Einstein Galactic plane survey conducted by \citet{1984ApJ...278..137H}. It was discovered with the Spacelab 2 X-Ray Telescope in 1985 August and detected by the ROSAT All-Sky Survey in 1990 \citep{1988MNRAS.232..551W,1990MNRAS.243...72S,1998A&AS..132..341M,2003ApJ...596.1220W,2016A&A...587A..61C}. The outbursts of SLX 1746-331 have been reported in 2003 \citep{2003ATel..143....1M}, 2007 \citep{2003ATel..143....1M}, and 2011 \citep{2011ATel.3098....1O}.  \citet{2015ApJ...805...87Y} estimated the distance of SLX 1746-331 to be about $10.81\pm3.52$ kpc using data from RXTE. \citet{2024ApJ...965L..22P} using the empirical mass-luminosity correlation of BHs estimates the mass of the black hole to be $5.5\pm3.6M_{\sun}$ based on NuSTAR, NICER, and \textit{Insight}-HXMT data during the 2023 outburst, and also gives the spin $a_{*}=0.85\pm0.03$ and inclination angle $i=53.0\pm0.5$ deg.

In this work, we aim to examine X-ray spectra observed by \textit{Insight}-HXMT during the 2023 outburst of SLX 1746-331 and to measure its spin. We jointly use a continuum-fitting model \texttt{kerrbb} and a reflection model \texttt{relxillcp} to fit the spectra. Observations, data selection, and reduction procedures are described in Section 2. Analysis methods and results are presented in Section 3. A discussion of the fitting results is written in Section 4. Section 5 is a summary of our work.

\section{Observations and data reduction}

\begin{figure}
	\includegraphics[width=1.\columnwidth]{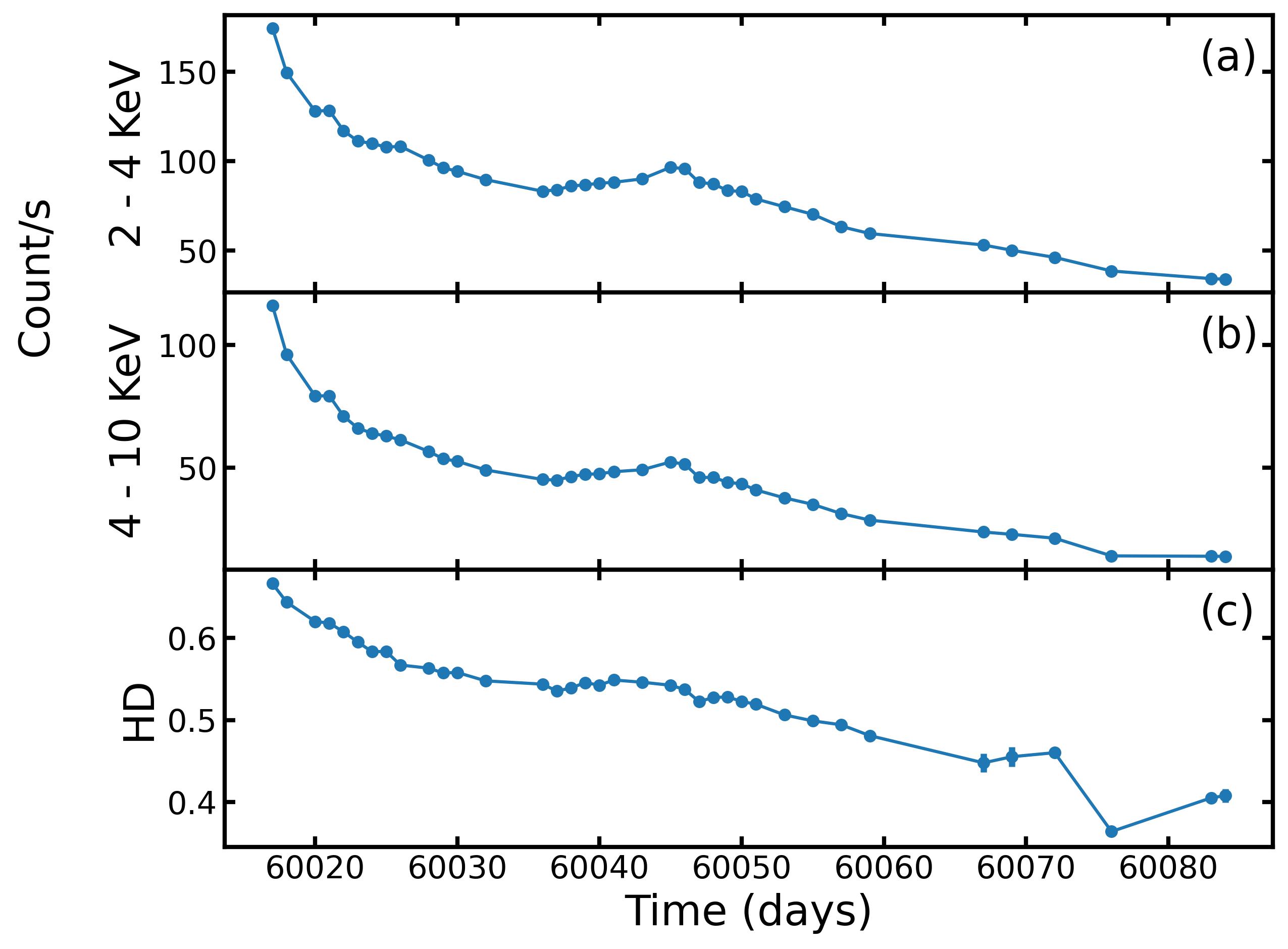}
    \caption{\textit{Insight}-HXMT observations light curve in the energy bands: (a) 2-4 keV and (b) 4-10 keV. Panel (c) is hardness ratio defined as the ratio of count rate between 4-10 keV and 2-4 keV.}
    \label{figure2}
\end{figure}

\begin{figure}
    \centering
	\includegraphics[width=1.0\columnwidth]{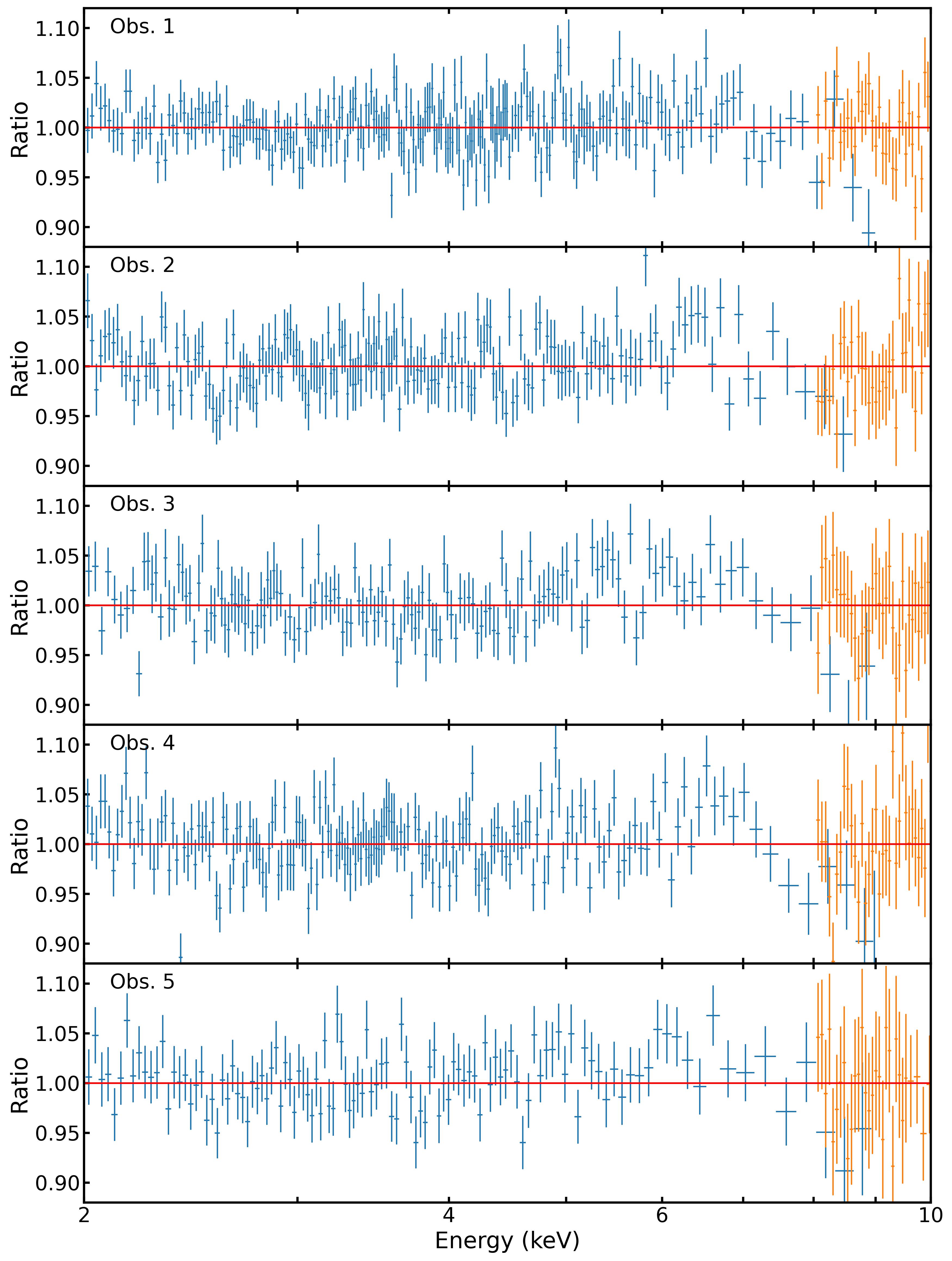}
    \caption{The data-to-model ratios of 5 selected observations in $2-10$ keV band. The spectra were fitted using the model \texttt{constant*tbabs*(diskbb+powerlaw)}. All observations show a weak iron line.}
    \label{ratio}
\end{figure}

\begin{figure*}
    \centering
	\includegraphics[width=1.3\columnwidth]{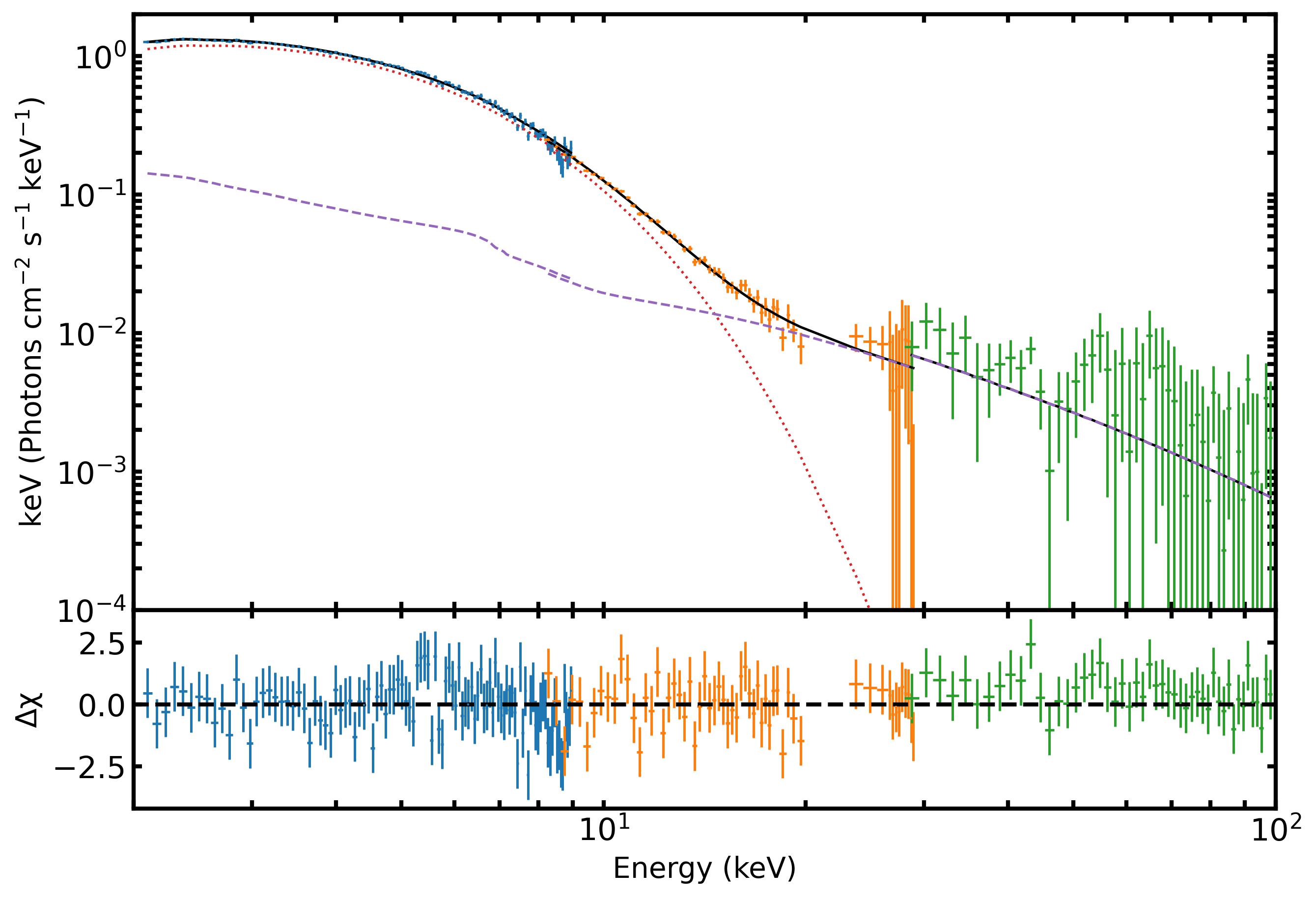}
    \caption{The result of spectral fitting to observation 3 with the model \texttt{constant*tbabs*(kerrbb+relxillcp)} and the $\Delta\chi$ of the fitting. The marks in blue are LE band data, the marks in orange are ME band data, and the marks in green are HE band data. The red dotted line is the \texttt{kerrbb} component, and the purple dashed line is the \texttt{relxillcp} component. The fitting returns a $\chi^2_\nu=0.95$ and the model fits the spectrum well.}
    \label{figure3}
\end{figure*}

The Hard X-ray Modulation Telescope \textit{Insight}-HXMT as China's first X-ray astronomy satellite \citep{2020SCPMA..6349502Z} is a large X-ray astronomical satellite with a broad energy band of 1-250 keV. To fulfill the requirements of the broadband spectra and fast variability observations, three payloads are configured onboard \textit{Insight}-HXMT: High Energy X-ray telescope (HE) for 20-250 keV band \citep{liu2020high}, Medium Energy X-ray telescope (ME) for 5-30 keV band \citep{cao2020medium}, and Low Energy X-ray telescope (LE) 1-15 keV band \citep{chen2020low}. Light curves and spectra were extracted using \textit{Insight}-HXMT Data Analysis Software (HXMTDAS) v2.05 following the standard procedure (also see processing details described in \citealt{WANG20211,chen2021relation}). In the data screening procedure, we use tasks $he/me/lepical$ to remove spike events caused by electronic systems and $he/me/legtigen$ to select good time interval (GTI) when the pointing offset angle $< 0.04^\circ$; the pointing direction above earth $> 10^\circ$; the geomagnetic cut-off rigidity > 8 GeV and the South Atlantic Anomaly (SAA) did not occur within 300 seconds. We used spectral analysis software Xspec v12.14.0 to study the spectra \citep{1996ASPC..101...17A}.

Our study is based on \textit{Insight}-HXMT observations of the 2023 outburst of SLX 1746-331 from 14 March 2023 (MJD 60017) to 19 May 2023 (MJD 60083). We collect all observations and plot the hardness-intensity diagram (HID) in Fig. \ref{figure1}. We fit all spectra by a simple model \texttt{constant*tbabs*(diskbb+powerlaw)} and select five observations that showed Fe K$\alpha$ lines to perform the spectral analysis in the BH spin study. The data-to-model ratios of these observations are plotted in Fig. \ref{ratio}. In Table \ref{table1}, we present the detailed information of these observations. In this work, we analyzed the spectra using 2-9 keV for LE \citep{chen2020low}, 8-29 keV for ME \citep{cao2020medium}, and 27-100 keV for HE \citep{liu2020high}. Data between 20 and 23 keV are ignored due to the calibration issues related to the silver line structure.

\section{analysis and results}

Figure \ref{figure2} presents the background-subtracted light curves from Insight-HXMT in 2-4 keV, 4-10 keV energy bands and the hardness ratios (ratio of count rate between 4-10 keV and 2-4 keV). The hardness-intensity diagram (HID) is shown in Fig. \ref{figure1}. The HID of the source shows that it evolved from high luminosities with hard spectra to low luminosities in a soft state and most observations are in the intermediate state.

To estimate the spin of the source, we fit the spectra with model \texttt{constant$*$tbabs*(kerrbb+relxillcp)}. Model \texttt{kerrbb} is a multi-temperature blackbody model for a thin, steady state, general relativistic accretion disk around a Kerr black hole \citep{2005ApJS..157..335L}. Model \texttt{relxillcp} is a model for relativistic reflection with the \texttt{nthcomp} model as the primary source spectrum \citep{2014ApJ...782...76G,2014MNRAS.444L.100D}. Besides, we tried to fit the spectra with the other reflection model \texttt{relxilllp}. The \texttt{relxilllp} model assumes that the corona is a point source located at a height above the central compact object. Most of the parameters in the \texttt{relxilllp} model are the same as in \texttt{relxill}. But instead of the emissivity index, \texttt{relxilllp} has two new parameters, $h$ and $\beta$, which are the height of the corona and the corona velocity. The model \texttt{relxilllp} gives similar results of inclination angle and spin to \texttt{relxillcp} (see Tables \ref{table2} \& \ref{table3} ).

\begin{figure}
	\includegraphics[width=\columnwidth]{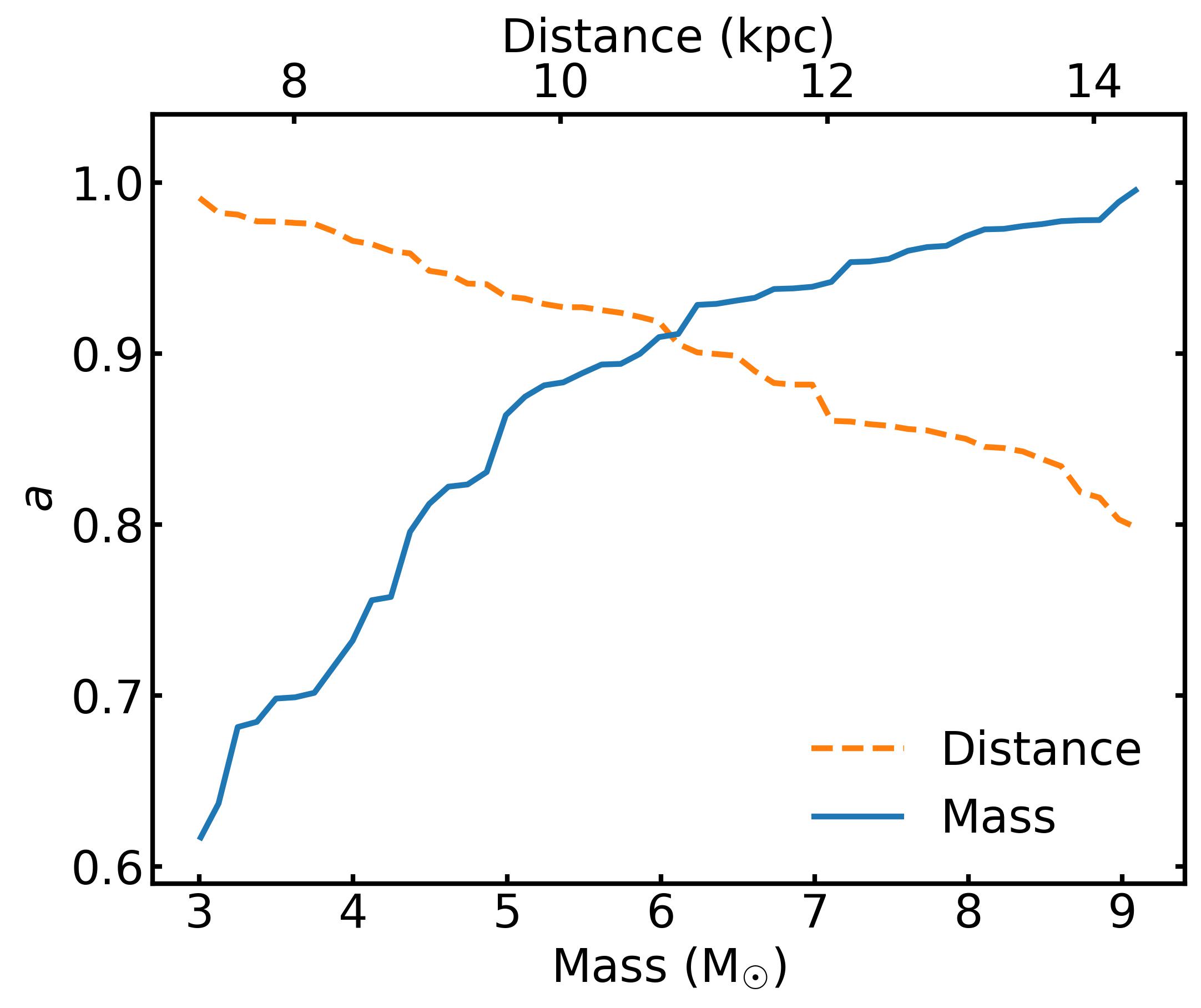}
    \caption{The correlation plots display the effect on the spin of varying mass and distance of the black hole. \textbf{Orange dashed line}: the spin versus the distance ($D=7.1-14.2$ kpc). \textbf{Blue solid line}: the spin versus the BH mass ($M=3.0-9.1~M_{\sun}$).}
    \label{figure4}
\end{figure}

Usually, if the normalization of LE is fixed to 1, the normalization of ME and HE should close to 1. However, there are minor differences between the calibration of the two detectors due to the effects of systematic errors \citep{2020JHEAp..27...64L}. The relative differences may change slightly during the fitting process. Model \texttt{constant} is used for coordinating calibration differences between detectors. In the fitting, we fixed the normalization of LE to 1, the normalizations of ME and HE are set free in the range of 0.85 to 1.15. Model \texttt{tbabs} fits the galactic absorption column density \citep{2000ApJ...542..914W} which is set to be free. The multi-temperature blackbody model \texttt{kerrbb} and the reflection model \texttt{relxillcp} are used to estimate the spin. In the model \texttt{kerrbb}, we set the black hole mass free, because the value of spin depends on the mass of the black hole (see discussion and Fig. \ref{figure4}) and the mass of SLX 1746-331 is not well constrained. The distance to the source is fixed to $D=10.81$ kpc \citep{2015ApJ...805...87Y}. The spin and inclination angle in \texttt{kerrbb} and \texttt{relxillcp} are linked. The normalization in \texttt{kerrbb} is fixed to 1. The rest of the parameters are set to their default values. 

\citet{2023ApJ...955...96P} found the disk flux and the inner disk temperature have a power law relation with a power law index around $3.980\pm0.004$ using data of \textit{Insight}-HXMT and NICER (see Fig. 4 in \citealt{2023ApJ...955...96P}), which suggests a thin disk and the soft state for the source \citep{2004MNRAS.347..885G,2011MNRAS.411..337D}. The five observations we selected have a luminosity $L\sim0.2-0.3L_{\rm{Edd}}$ ($L_{\rm{Edd}}$, Eddington luminosity) assuming a BH mass $\rm{M}=\rm{5.8\pm0.3~M_{\sun}}$ and a distance $D=10.81$ kpc. These indicate that the inner radius of the disk is near the ISCO for the five observations \citep{2010ApJ...718L.117S,2013MNRAS.431.3510S}. In the model \texttt{relxillcp}, we fix the inner radius of the accretion disk $R_{in}=-1$ (at the ISCO). The emissivity index of the disk is frozen to 3. We fix the iron abundance $\rm{A_{Fe}}=1$ and set the density of accretion disk $\rm{logN}$ free. The electron temperature in the corona (kTe) is fixed to 300 keV. The free parameters in the model \texttt{relxillcp} are the power-law index of the primary source spectrum ($\Gamma$), the density of accretion disk ($\rm{logN}$), the ionization of the accretion disk ($\log\xi$), reflection fraction parameter (${R_{f}}$), the black hole mass ($M$), the spin of the BH ($a_{*}$), and the inclination of the disk ($i$).

\begin{table*}
\renewcommand{\arraystretch}{1.5}
\setlength{\tabcolsep}{4mm}
    \centering
    \caption{Fitting results of \textit{Insight}-HXMT Observations of SLX 1746-331 in the 2023 outburst with model \texttt{constant*tbabs(kerrbb+relxillcp)}. The letter P indicates that the error of the parameter was pegged at the upper or lower boundary, and the letter f indicates the parameter was fixed at the given value. All errors were calculated with 90 percent confidence level.}
    \begin{threeparttable}[b]
    \begin{tabular}{c c c c c c c}
       \hline
       Model    &   Parameters           &   Obs. 1    &   Obs. 2    &   Obs. 3    &   Obs. 4    &   Obs. 5   \\
       \hline
       Tbabs    &   $\rm{N_H(\times10^{22}~cm^{-2})}$  & $1.04^{+0.02}_{-0.02}$ & $1.02^{+0.01}_{-0.01}$ & $1.08^{+0.11}_{-0.10}$ & $0.98^{+0.05}_{-0.06}$ & $1.03^{+0.03}_{-0.04}$ \\
       kerrbb   &    $\rm{M}$ $(\rm{M}_{\sun})$ & $6.1^{+0.1}_{-0.1}$ & $6.0^{+0.1}_{-0.1}$ & $6.0^{+0.3}_{-0.3}$ & $5.6^{+0.1}_{-0.2}$ & $5.9^{+0.1}_{-0.1}$ \\
                &   $\dot{\rm{M}}$ $(\times10^{18}~g{\hspace{0.1cm}}s^{-1})$ &  $1.69^{+0.01}_{-0.01}$      & $1.46^{+0.02}_{-0.02}$  &  $1.44^{+0.07}_{-0.07}$   &   $1.24^{+0.02}_{-0.03}$          &  $1.20^{+0.02}_{-0.02}$     \\
       relxillcp &    $i$ (deg)           &  $61.9^{+0.3}_{-0.3}$           & $61.8^{+0.5}_{-0.4}$       &  $60.5^{+1.1}_{-1.4}$        &   $60.0^{+0.8}_{-1.2}$       &   $60.6^{+0.7}_{-0.7}$      \\
                 &    $a_{*}$             &  $0.902^{+0.001}_{-0.001}$      & $0.903^{+0.002}_{-0.002}$  &  $0.907^{+0.012}_{-0.015}$   &   $0.902^{+0.006}_{-0.004}$  &  $0.903^{+0.003}_{-0.004}$  \\
                 &    $\Gamma$            &  $1.55^{+0.07}_{-0.06}$         & $1.73^{+0.17}_{-0.12}$     &  $2.01^{+0.01}_{-0.01}$      &   $1.63^{+0.03}_{-0.04}$     &  $2.13^{+0.05}_{-0.04}$     \\
                 &    $\rm{log}\xi$ (erg cm s$^{-1}$)       &  $3.03^{+0.02}_{-0.03}$         & $2.68^{+0.06}_{-0.04}$     &  $2.69^{+0.05}_{-0.05}$      &   $2.68^{+0.09}_{-0.06}$     &   $2.68^{+0.04}_{-0.04}$    \\
                 &    $\rm{log}N$ $({cm^{-3})}$    &  $17.1^{+0.1}_{-0.2}$  & $16.7^{+1.0}_{-1.5}$       &  $15.7^{+0.4}_{-0.5}$        &   $15.9^{+0.6}_{-0.4}$       &  $15.7^{+0.5}_{-0.6}$       \\
                 &    $kT_e$ (keV) &             \multicolumn{5}{c}{$300$ (f)}                  \\
                 &    $R_f$               &  $1.50^{+0.02}_{-0.02}$         & $2.15^{+0.02}_{-0.01}$     &  $3.35^{+0.04}_{-0.05}$      &   $1.67^{+0.02}_{-0.03}$     &  $1.96^{+0.02}_{-0.03}$     \\
                 &    $\rm{Norm_{relxillcp}}$ ($\times10^{-3}$) &  $1.99^{+0.10}_{-0.13}$ & $1.12^{+0.01}_{-0.03}$ &  $1.10^{+0.01}_{-0.01}$ &  $1.36^{+0.02}_{-0.02}$ &  $1.33^{+0.12}_{-0.07}$     \\
                 &    $\chi^2/d.o.f$        &  314/248      & 316/248            &  250/248           &   235/248        &  226/248         \\
       \hline
    \end{tabular}
    \end{threeparttable}
    \label{table2}
\end{table*}

\begin{table*}
\renewcommand{\arraystretch}{1.5}
\setlength{\tabcolsep}{4mm}
    \centering
    \caption{Fitting results of \textit{Insight}-HXMT Observations of SLX 1746-331 in the 2023 outburst with model \texttt{constant*tbabs(kerrbb+relxilllp)}. The letter P indicates that the error of the parameter was pegged at the upper or lower boundary, and the letter f indicates the parameter was fixed at the given value. All errors were calculated with 90 percent confidence level.}
    \begin{threeparttable}[b]
    \begin{tabular}{c c c c c c c}
       \hline
       Model    &   Parameters           &   Obs. 1    &   Obs. 2    &   Obs. 3    &   Obs. 4    &   Obs. 5   \\
       \hline
       Tbabs    &   $\rm{N_H(\times10^{22}~cm^{-2})}$  & $1.24^{+0.01}_{-0.02}$ & $1.18^{+0.09}_{-0.07}$ & $1.21^{+0.01}_{-0.01}$ & $1.34^{+0.01}_{-0.01}$ & $1.30^{+0.04}_{-0.04}$ \\
       kerrbb   &    $\rm{M}$ $(\rm{M}_{\sun})$ & $5.5^{+0.1}_{-0.1}$ & $5.8^{+0.1}_{-0.1}$ & $5.4^{+0.1}_{-0.1}$ & $5.4^{+0.1}_{-0.1}$ & $6.0^{+0.2}_{-0.2}$ \\   
                 & $\dot{\rm{M}}$ $(\times10^{18}~g{\hspace{0.1cm}}s^{-1})$ & $1.73^{+0.02}_{-0.01}$ & $1.39^{+0.04}_{-0.03}$ & $1.53^{+0.01}_{-0.02}$ & $1.28^{+0.01}_{-0.01}$ & $1.19^{+0.04}_{-0.04}$ \\
       relxilllp &      $h$ (${R_g}$)  &  $3.1^{+0.1}_{-0.1}$         & $3.0^{+0.2}_{-P}$     &  $3.0^{+0.1}_{-P}$     &   $3.0^{+0.1}_{-P}$     &  $3.3^{+0.1}_{-0.1}$  \\
                 &    $i$ (deg)           &  $60.0^{+0.3}_{-0.2}$           & $60.0^{+0.2}_{-0.3}$       &  $59.1^{+0.3}_{-0.2}$       &   $59.2^{+0.1}_{-0.5}$       &  $60.7^{+0.4}_{-0.4}$      \\
                 &      $a_{*}$               &  $0.881^{+0.004}_{-0.004}$      & $0.906^{+0.003}_{-0.005}$  &  $0.881^{+0.001}_{-0.001}$  &   $0.880^{+0.002}_{-0.002}$  &  $0.901^{+0.008}_{-0.010}$ \\
                 &    $\Gamma$            &  $2.17^{+0.02}_{-0.03}$         & $2.27^{+0.02}_{-0.01}$     &  $2.23^{+0.05}_{-0.04}$     &   $2.42^{+0.03}_{-0.02}$     &  $2.50^{+0.04}_{-0.04}$    \\
                 &    $\rm{log}\xi$ (erg cm s$^{-1}$)      &  $2.98^{+0.02}_{-0.02}$         & $2.97^{+0.04}_{-0.03}$     &  $2.60^{+0.01}_{-0.01}$     &   $2.80^{+0.05}_{-0.02}$     &  $2.99^{+0.15}_{-0.11}$    \\
                 &    $\rm{A_{Fe}}$       &  $4.8^{+0.1}_{-0.1}$         & $2.8^{+0.2}_{-0.2}$     &  $8.5^{+0.2}_{-0.3}$        &   $10.0^{P}_{-0.1}$        &  $5.8^{+0.1}_{-0.1}$  \\
                 &    $kT_e$ (keV)        &             \multicolumn{5}{c}{$300$ (f)}              \\
                 &    $R_f$               &  $1.13^{+0.01}_{-0.01}$      & $2.53^{+0.12}_{-0.17}$  &  $2.70^{+0.3}_{-0.7}$   &   $1.80^{+0.01}_{-0.02}$ &  $1.11^{+0.02}_{-0.02}$    \\
                 &    $\rm{Norm_{relxilllp}}$ ($\times10^{-2}$) &  $2.21^{+0.03}_{-0.03}$ & $1.29^{+0.01}_{-0.01}$ & $1.89^{+0.01}_{-0.03}$ & $3.87^{+0.03}_{-0.01}$ &  $3.29^{+0.01}_{-0.01}$    \\
                 &    $\chi^2/d.o.f$        &  326/247             & 314/247           &  249/247           &   243/247         &  222/247           \\
       \hline
    \end{tabular}
    \end{threeparttable}
    \label{table3}
\end{table*}

Figure \ref{figure3} shows an example of the Insight-HXMT spectral fitting result from 2 - 100 keV. The best-fitting spectral parameters are listed together in Table \ref{table2} and Table \ref{table3} for \texttt{relxillcp} and \texttt{relxilllp} respectively, two models return similar values of the BH mass, spin and disk inclination. The five observations are similar without showing significant differences in the parameter values. However, the BH mass, the spin, the disk inclination angle, and the disk density should be the same in these observations, since their values cannot vary on a time scale of a few days. For this reason, we carried out a joint spectral fit for all five observations using model \texttt{constant*tbabs*(kerrbb+relxillcp)}. In the joint fit, the spin parameter, inclination angle, and disk density were tied, the electron temperature in the corona (kTe) was fixed to 300 keV, whereas other parameters were allowed to vary among the observations. The best-fit model parameters are listed in Table \ref{table4}. The joint fit yields the spin $a_{*}=0.88^{+0.1}_{-0.2}$, disk inclination $i=63.7^{+1.3}_{-1.0}$ deg, and the black hole mass $\rm{M}=\rm{5.8\pm0.3~M_{\sun}}$ (all at 90 percent statistical confidence).


\section{discussion}

SLX 1746-331 showed a different behavior from other BH candidates (like Q-diagram) in the HID: it evolved from high luminosities with relative hard spectra to low luminosities in relative soft state, and a similar trend is observed in the BH X-ray binary 4U 1543-47 \citep{2024MNRAS.527..238C}. In a classical successful outburst, the transient low-mass BH X-ray binary will transfer from the quiescent state to the low/hard state (LHS) and then through the intermediate-state (IMS) to the high/soft state (HSS) and back into the LHS as the accretion rate increases. The trace of a typical BH X-ray binary in HID is a counterclockwise "q" shape \citep{2005Ap&SS.300..107H,2019NewAR..8501524I}. However, not all outbursts follow a complete "q" shape. In some sources, the accretion rate is so low that the outburst does not reach the HSS but only evolves to the low/hard or intermediate state before ending. These outbursts are known as "failed outbursts" and account for approximately 38\% of all outbursts \citep{2004NewA....9..249B,2009MNRAS.398.1194C,2016ApJS..222...15T}. In contrast to the absence of the LHS, there is a class of outbursts that do not have an LHS at the beginning of the outburst, or whose LHS is not observed. Such outbursts are relatively rare, as seen in sources like 4U 1630-472 \citep{2020MNRAS.497.1197B} and MAXI J0637-430 \citep{2022MNRAS.514.5238M}. The 2023 outburst of SLX 1746-331 didn't follow a complete "q" shape in HID. Therefore, we cannot determine its spectral state via HID. However, \citet{2023ApJ...955...96P} studied the relation between the disk flux and the inner disk temperature, and they concluded that SLX 1746-331 stayed mostly in the soft state.

If the X-ray source is point-like and at height $h$ on the disk axis, the irradiation of the disk at large radius in the absence of any relativistic effect is proportional to $(r^2+h^2)^{-3/2}\propto r^{-3}$ \citep{1997ApJ...488..109R}. Therefore, the emissivity profile with index $q$ is generally assumed to be 3. In the fitting, we find that the value of spin is insensitive to the value of emissivity index and we fix the emissivity index of the disk to $q=3$.

So far, no prevailing explanation has been preferred for the black hole systems to be iron-rich. Nonetheless, supersolar prediction for the Fe abundance has been reported in some stellar-mass black hole binaries such as GX 339-4 \citep{2015ApJ...813...84G}, V404 Cyg \citep{2017ApJ...839..110W} and Cyg X-1 \citep{2015ApJ...808....9P}. A similar trend is found in AGNs as well, the iron of Seyfert galaxy 1H0707-495 is overabundant by a factor of 10-20 \citep{2009Natur.459..540F,2012MNRAS.422.1914D}. \citet{2018ASPC..515..282G} collected the reports of iron abundance obtained by reflection models for 13 AGNs and 9 BHBs, finding that iron abundance has a trend of a few times over the solar value in both AGNs and BHBs. Supersolar abundances are unlikely realistic since metal enrichment mechanisms in these two types of systems are expected to be very different, and the most likely explanation for the supersolar iron abundance results is the prediction of a relatively low disk density \citep{2018ASPC..515..282G}. For this reason, we fix the iron abundance $\rm{A_{Fe}}=1$ and set the density of accretion disk $\rm{logN}$ free in model \texttt{relxillcp}. Our joint fit returns a density $\rm{logN}=15.7^{+0.2}_{-0.3}$ cm$^{-3}$. In model \texttt{relxilllp}, the disk density is assumed to be $10^{15}$ cm$^{-3}$ and is not a free parameter. Therefore, we set the iron abundance $\rm{A_{Fe}}$ free in the fitting. Predictions based on the standard $\alpha$-disk model \citep{1973A&A....24..337S} and more sophisticated 3D magneto-hydrodynamic (MHD) simulations \citep{2010ApJ...711..959N,2013ApJ...769..156S} suggest densities in black hole accretion disks orders of magnitude larger than $\rm{N}\sim 10^{15}$cm$^{-3}$, our fitting result is consistent with the prediction. Nevertheless, the supersolar iron abundance in AGN may be a real effect. \citet{2012ApJ...755...88R} showed that radiative levitation of iron ions in the innermost regions of radiation-dominated AGN disks can enhance the photospheric abundance of iron.

The ionization parameter is defined by:
\begin{equation}
    \xi=4\pi F_\textrm{X}(r)/n_\textrm{e}(r),
\end{equation}
where $F_\textrm{X}(r)$ is the X-ray flux of the irradiation and $n_\textrm{e}(r)$ is the electron density of the disk at radius $r$ \citep{2000PASP..112.1145F}. The strong radius dependence of $F_\textrm{X}(r)$ will lead to the radial decrease of the disk ionization for any reasonable density profile of the disk \citep{2012A&A...545A.106S}. In the reflection models \texttt{relxillcp} and \texttt{relxilllp}, the ionization is assumed to be constant across the disk, its value ranges from 0 (neutral) to 4.7 (heavily ionized) and is sensitive to both the disk structure and the coronal illumination \citep{2011ApJ...734..112B,2014ApJ...782...76G}. The ignorance of the ionization gradient can lead to an increase in the emissivity index \citep{2012A&A...545A.106S,2019MNRAS.485..239K}. However, some models with non-trivial ionization gradients provide very similar results to standard models with constant ionization \citep{2021PhRvD.103j3023A,2022MNRAS.517.5721M}, the impact of ionization gradient is modest and models with constant electron density and ionization parameter are probably sufficient in most cases \citep{2022MNRAS.517.5721M}. Therefore, we didn't use the models that take the ionization gradient into consideration during the fitting. Our fittings return an ionization parameter $\rm{log}\xi\sim2.4-3.1$.

The model \texttt{kerrbb} relies on the system parameters of mass, distance, and inclination of the source. The spin parameter measured by model \texttt{kerrbb} decreases as the mass decreases or the distance increases (see Figure 5 in \citealt{2021ApJ...916..108Z}). In this work, we used a joint model of \texttt{kerrbb} and \texttt{relxillcp} to estimate the spin. To study the influence of black hole mass and distance on the joint model, we fit the Obs. 3 with masses ranging between $3.0M_{\sun}<M<9.1M_{\sun}$ \citep[stellar-mass black holes are expected to have masses around 3 to 100 $M_{\sun}$, ][]{2003ApJ...596..437C} and distance of $7.1~\rm{kpc}<D<14.2~\rm{kpc}$, the inclination range is set based on the best-fitting value and errors in Table \ref{table2}. The results are plotted in Figure \ref{figure4}, the spin parameter increases as the mass increases or distance decreases. The influence of the variation of mass and distance on the spin in the joint model is similar to the result of the single model \texttt{kerrbb}.
Our results indicate that the spin measured by the joint model depends on the mass of the black hole. Since the mass of SLX 1746-331 is not well constrained, we set the mass of the black hole free during the fitting. The joint fit gives the black hole mass $\rm{M}=\rm{5.8\pm0.3~M_{\sun}}$, which is consistent with the estimation of \citet{2024ApJ...965L..22P}. A more accurate measurement of the spin can be obtained if the mass can be constrained better in future work.

The spin we estimated in the work is similar to the value given by \citet{2024ApJ...965L..22P}, but the inclination we obtained is about 10$^{\circ}$ larger than their result. \citet{2024ApJ...965L..22P} used \texttt{ezdiskbb} and \texttt{relxillNS} \citep{2022ApJ...926...13G} to fit the spectra. In the model \texttt{relxillNS}, the input spectrum is a black body with temperature $kT_{bb}$, and it can model the thermal radiation of a neutron star incident on the accretion disk. When using \texttt{relxillNS} for a black hole X-ray binary, it approximates the returning radiation of the thermal spectrum of the disk \citep{2020ApJ...892...47C,2021ApJ...921..155L,2021ApJ...906...11W,2021ApJ...909..146C}. The difference between these models may lead to a difference in the estimation of the angle. In addition, the density of the accretion disk can affect the value of the inclination angle \citep{2015ApJ...813...84G,2018ApJ...855....3T,2019MNRAS.484.1972J}. The blue wing of the broad Fe line principally determines the inclination angle estimation via the X-ray reflection fitting method. The high-density model will lead to increasing soft X-ray flux. Recent reflection analyses of Cyg X-1 by \citet{2018ApJ...855....3T}, GX 339-4 by \citet{2015ApJ...813...84G} and \citet{2019MNRAS.484.1972J} suggest that reflection models that underestimate the density of disk can introduce systematic changes of order 10 deg in the inclination angle. In \texttt{relxillNS}, the disk density is assumed to be $10^{15}$cm$^{-3}$ and is not variable. However, we set the disk density free in our fitting and obtain $\rm{N}\sim10^{16}$cm$^{-3}$. The underestimation of the disk density in the model \texttt{relxillNS} may lead to a deviation in the estimation of the inclination angle.

In addition to \texttt{relxillcp}, which doesn't assume any geometry of corona and take any relativistic boosting effects into account \citep{2016AN....337..362D}, we also tried to fit the spectra with \texttt{relxilllp} assuming that the corona is a point source located at a height $h$ above the accretion disk along the spin axis of the black hole. The height is a key parameter for the ionization and the disk reflection. In \texttt{relxilllp}, the illumination source of the reflection is assumed to be the corona, which is described as a power-law with a high-energy cut-off \citep{2013MNRAS.430.1694D}. The best-fitting spin and inclination in the two models are similar. The corona is located close to the black hole, at the height of $h\sim3~R_g$ (the theoretical value of corona height is $3-100~R_g$ \citet{2017MNRAS.472L..60B}).

\section{Conclusion}

In this work, we have measured the spin and inclination angle of SLX 1746-331 by modeling its black body and reflection components in five observations observed by \textit{Insight}-HXMT during the 2023 outburst. The spectra consist of three different components: galactic absorption, thermal emission from the accretion disk, and reflection emission. We fit the spectra separately and jointly fit all five spectra for a consistent result of spin. We used model \texttt{relxillcp} to fit the reflection component and model \texttt{kerrbb} to fit the thermal emission from the accretion disk.

According to the results of joint-fitting, we constrain the spin $a_{*}=0.88^{+0.1}_{-0.2}$, the black hole mass $\rm{M}=\rm{5.8\pm0.3~M_{\sun}}$ and disk inclination $i=63.7^{+1.3}_{-1.0}$ deg. The spin is similar to the estimate conducted by \citet{2024ApJ...965L..22P}, the BH mass is also consistent with the previous value; however, the inclination is about 10$^{\circ}$ larger compared to their result. The difference may be due to the difference between models or underestimation of the disk density. Besides, we study the effects of mass and distance of black hole on the estimation of spin. The spin parameter increases as the mass increases or distance decreases. Therefore, a more accurate measurement of the black mass and distance is needed to have a better constraint on the spin. 

\begin{table*}
\renewcommand{\arraystretch}{1.5}
\setlength{\tabcolsep}{4mm}
    \centering
    \caption{Fitting results of \textit{Insight}-HXMT Observations of SLX 1746-331 in the 2023 outburst with model \texttt{constant*tbabs(kerrbb+relxillcp)}. The letter P indicates that the error of the parameter was pegged at the upper or lower boundary, and the letter f indicates the parameter was fixed at the given value. All errors were calculated with 90 percent confidence level.}
    \begin{threeparttable}[b]
    \begin{tabular}{c c c c c c c}
       \hline
       Model    &   Parameters          &   Obs. 1    &   Obs. 2    &   Obs. 3    &   Obs. 4    &   Obs. 5   \\
       \hline
       Tbabs    &   $\rm{N_H(\times10^{22}~cm^{-2})}$  &            \multicolumn{5}{c}{$1.00^{+0.05}_{-0.04}$}                \\
       kerrbb   &    $\rm{M}$ $(\rm{M}_{\sun})$    &             \multicolumn{5}{c}{$5.8\pm0.3$}     \\
                 & $\dot{\rm{M}}$ $(\times10^{18}~g{\hspace{0.1cm}}s^{-1})$ & $1.87^{+0.05}_{-0.03}$ & $1.64^{+0.05}_{-0.03}$ & $1.65^{+0.05}_{-0.03}$ & $1.42^{+0.05}_{-0.03}$ & $1.35^{+0.04}_{-0.03}$ \\
       relxillcp &    $i$ (deg)           &           \multicolumn{5}{c}{$63.7^{+1.3}_{-1.0}$}                \\
                 &      $a_{*}$               &           \multicolumn{5}{c}{$0.88^{+0.01}_{-0.02}$}                \\
                 &    $\Gamma$            &  $1.53^{+0.05}_{-0.04}$         & $1.66^{+0.05}_{-0.06}$     &  $1.90^{+0.02}_{-0.04}$      &   $1.60^{+0.05}_{-0.05}$     &  $2.19^{+0.06}_{-0.07}$     \\
                 &    $\rm{log}\xi$ (erg cm s$^{-1}$)      &  $3.09^{+0.11}_{-0.09}$         & $2.66^{+0.12}_{-0.14}$     &  $2.69^{+0.03}_{-0.07}$      &   $2.36^{+0.04}_{-0.11}$     &   $2.68^{+0.11}_{-0.20}$    \\
                 &    $\rm{log}N$ $({\rm cm^{-3})}$ &     \multicolumn{5}{c}{$15.7^{+1.2}_{-0.6}$}                \\
                 &    $kT_e$ (keV)        &     \multicolumn{5}{c}{$300$ (f)}                \\
                 &    $R_f$               &  $1.54^{+0.06}_{-0.05}$      & $2.07^{+0.03}_{-0.03}$  &  $3.83^{+0.11}_{-0.18}$   &   $1.26^{+0.03}_{-0.05}$     &  $1.99^{+0.04}_{-0.03}$     \\
                 &    $\rm{Norm_{relxillcp}}$ ($\times10^{-3}$) &  $2.10^{+0.11}_{-0.11}$ & $1.30^{+0.03}_{-0.02}$ &  $0.92^{+0.02}_{-0.02}$ &  $1.72^{+0.04}_{-0.04}$ &  $1.51^{+0.03}_{-0.05}$     \\
                 &    $\chi^2/d.o.f$        &     \multicolumn{5}{c}{$1347/1258$}                \\
       \hline
    \end{tabular}
    \end{threeparttable}
    \label{table4}
\end{table*}

\section*{Acknowledgements}
We are grateful to the referee for the fruitful comments. This work is supported by the National Key Research and Development Program of China (Grants No. 2021YFA0718503 and 2023YFA1607901), the NSFC (12133007). This work has made use of data from the \textit{Insight-}HXMT mission, a project funded by the China National Space Administration (CNSA) and the Chinese Academy of Sciences (CAS).

\bibliographystyle{aasjournal}



\end{document}